\documentclass[aps,prb,showpacs,superscriptaddress,twocolumn,amsmath]{revtex4}
\usepackage{txfonts}
\usepackage[pdftex]{graphicx}
\usepackage{bm}
\usepackage{overpic}
\usepackage{bbold}
\def\rnum#1{\expandafter{\romannumeral #1}} 
\def\Rnum#1{\uppercase\expandafter{\romannumeral #1}}

\newfont{\bg}{cmr10 scaled\magstep4}

\newcommand{\bigzerou}{\smash{\lower1.8ex\hbox{\bg 0}}}

\begin{document}

\title{Rotation of optically bound particle assembly due to scattering induced spin-orbit coupling of light}

\author{Yukihiro Tao}
\affiliation{Department of Material Engineering Science,
Graduate School of Engineering Science, Osaka University,
1-3 Machikaneyama, Toyonaka, Osaka 560-8531, Japan}

\author{Tomohiro Yokoyama}
\email[E-mail me at: ]{tomohiro.yokoyama@mp.es.osaka-u.ac.jp}
\affiliation{Department of Material Engineering Science,
Graduate School of Engineering Science, Osaka University,
1-3 Machikaneyama, Toyonaka, Osaka 560-8531, Japan}

\author{Hajime Ishihara}
\affiliation{Department of Material Engineering Science,
Graduate School of Engineering Science, Osaka University,
1-3 Machikaneyama, Toyonaka, Osaka 560-8531, Japan}
\affiliation{Department of Physics and Electronics,
Graduate school of Engineering, Osaka Prefecture University,
1-1 Gakuen-cho, Naka-ku, Sakai, Osaka 599-8531, Japan}
\affiliation{Center for Quantum Information and Quantum Biology,
Institute for Open and Transdisciplinary Research Initiatives, Osaka University,
1-3 Machikaneyama, Toyonaka, Osaka 560-8531, Japan}

\date{\today}

\begin{abstract}
The optical binding of many particles has great potential to achieve the wide-area formation of a ``crystal'' of small materials.
Unlike conventional optical binding, where the whole assembly of targeted particles is irradiated with light,
if one can indirectly manipulate remote particles using a single trapped particle through optical binding,
the degrees of freedom to create ordered structures will be greatly enhanced.
In this Letter, we theoretically investigate the dynamics of the assembly of gold nanoparticles
that is manipulated using a single particle trapped by a focused laser.
As a result, we demonstrate that the spin--orbit coupling and angular momentum generation of light via scattering induce
the assembly and rotational motion of particles through indirect optical force.
This result opens the possibility of creating ordered structures with a wide area and manipulating them,
controlling local properties using scanning laser beams.
\end{abstract}

\maketitle

\section{Introduction}
Optical manipulation is a non-contact method to capture various micro-scale objects~\cite{Ashkin1986},
such as metals, semiconductors, dielectrics, organic materials, and living cells~\cite{HZhang2008}.
Due to this variety of trappable objects, laser trapping has been developed for a wide range of research fields~\cite{TLi2010,Fazal2011}.
One significant development in optical manipulation is the trapping of multiple particles.
For instance, a holographical technique can form various chains of microparticles~\cite{Curtis2002,Grier2006}.
On a glass substrate, total reflection can provide a trapping force over a wide area~\cite{Mellor2006cpc,Mellor2006oe,Taylor2008}.
In addition to such design of incident light, micro-scale fabrication has also achieved the trapping of many particles,
e.g., plasmonic structures~\cite{Righini2007,YPang2012} and photonic crystals~\cite{Rahmani2006,AHJYang2009,Jaquay2013}.
The formation of an ordered monolayer of particles at a liquid--liquid interface has also been reported~\cite{Aveyard2002,BJPark2008}.

Optical binding is a key concept for the optical manipulation of many particles~\cite{Depasse1994,Forbes2020}.
Optically induced polarizations cause attractive or repulsive forces between the particles,
which results in an ordering of the particles~\cite{Demergis2012}.
Yan et al. investigated the formation of a nanoparticle array under wide-area laser irradiation with circular~\cite{FHan2018}
and linear polarizations~\cite{ZYan2013acsn,ZYan2014}, where the ordering of particles depended on the type of polarization.
The spin angular momentum (SAM) of circular polarization gives a torque to the array,
although the mechanism of this torque transfer is still unclear.

Recent reports by \ Kudo {\it et al.}~\cite{Kudo2016,SFWang2016,Kudo2018} demonstrated new possibilities of
optical manipulation and optical binding.
They examined the trapping of many nanoparticles by a tightly focused single laser,
where trapped particles showed tetragonal or hexagonal ordered arrays depending on the polarization.
These arrays were beyond the irradiation area, like a crystal growing~\cite{Sugiyama2007}.
Moreover, outside of the focal area, polystyrene particles form additional horns~\cite{Kudo2016} and
gold particles show revolution and swarming dynamics~\cite{Kudo2018}.
This implies that focal irradiation and scattering fields from the particles cause self-organized and indirect optical binding,
which is in significant contrast with conventional optical binding with a wide area of irradiation.
The mechanism of such indirect binding of multiple particles is unknown at present.
Its elucidation will lead to an unconventional scheme for creating and manipulating wide-area ordered structures, implementing
finely controlled local properties with scanning beams and rich extension of optical binding by
a combination of multiple focused lasers.

In this Letter, we theoretically investigates the mechanism of the optical binding and dynamics of nanoparticles due to
indirect optical force under the irradiation of a single, tightly focused, and circularly polarized laser.
Considering gold nanoparticles, we numerically demonstrate the revolution of surrounding particles.
The binding position is determined by the field intensity,
whereas the revolution is explained by a ``spin--orbit (SO) coupling'' of light.
An analysis based on the SAM and orbital angular momentum (OAM) of light elucidates three factors:
large scattering and suppressed absorption cross-sections, conversion of SAM to OAM by a single particle,
and imbalance of positive and negative OAM generation.
The first and second factors are determined by the individual particle's properties, which accelerates the revolution.
The revolution should relate to a vortical flow of the momentum of light.
The optical current discussed by Berry~\cite{Berry2009} describes such momentum flow of light.
The interference between incident and scattered field clearly exhibits a vortical structure of
the optical current, as shown in Appendix \ref{appen:ocof}.
The last factor is affected by the particle configurations as well as properties.
Our results imply that the imbalance of OAM will reveal the revolution direction.

\section{Model}
\begin{figure}[t]
\includegraphics[width=1.0\linewidth]{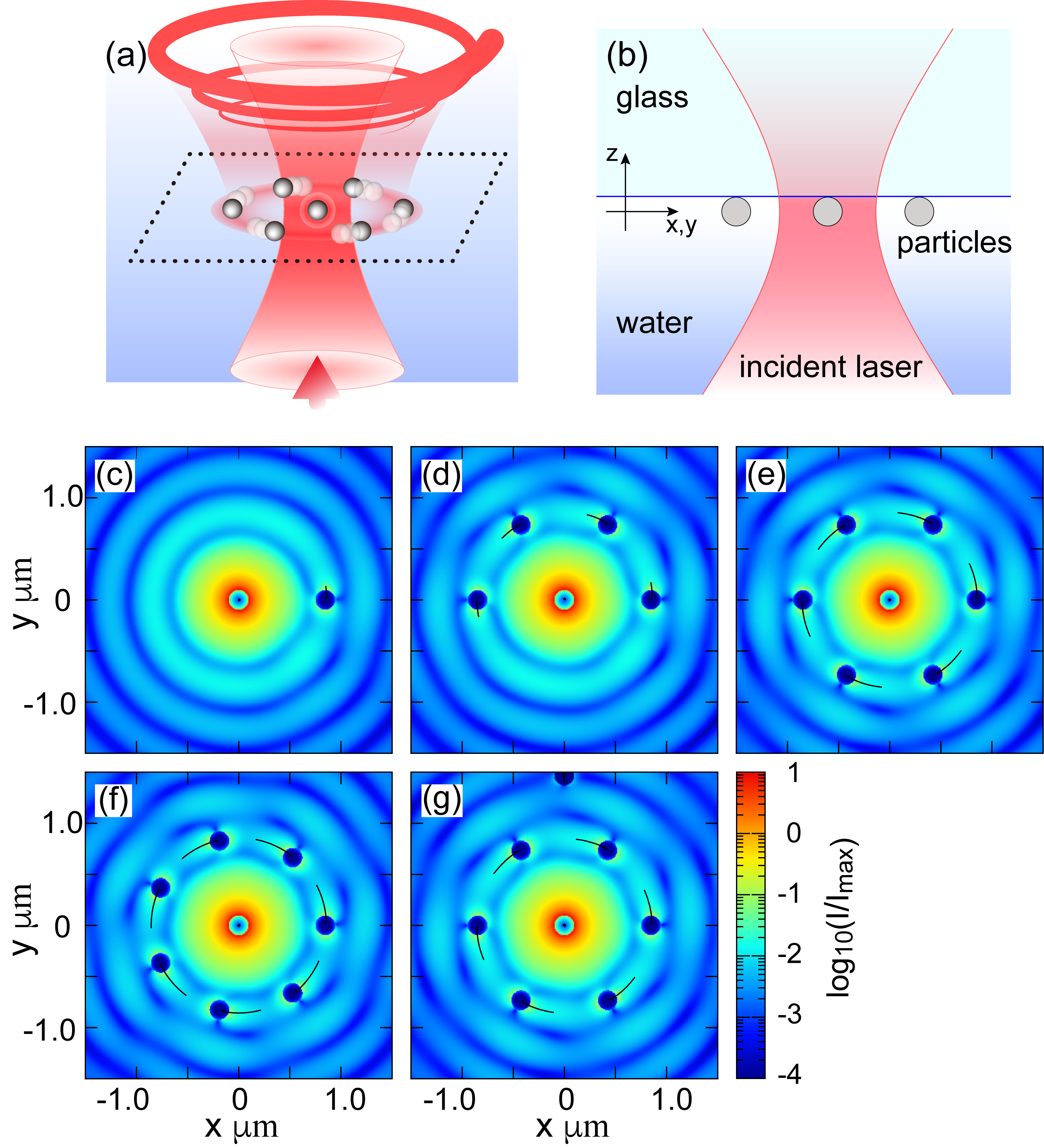}
\caption{
(a) Schematic view of optical trapping and binding by a single focused laser
accompanying by the generation of OAM, and (b) model of the numerical simulation
for $d=200\, \mathrm{nm}$ diameter Au particles in water.
(c)-(g) Intensity profiles of total electric field with $N_{\rm p} = 2$ (c), $5$ (d), $7$ (e), and $8$ (f,g) on the $z=0$ plane.
The incident laser is tightly focused with circular polarization.
The focal point is at the origin.
Black lines in the panels indicate trajectories of particles in a finite period.
}
\label{fig:OptBind}
\end{figure}

Considering the experiment reported in Ref.~\cite{Kudo2018}, we assume the following model and conditions.
We consider nanoscale spherical gold particles (refractive index: $n_{\rm Au} \simeq 0.258 + 6.97 i$) in
a water solvent ($n_{\rm w} \simeq 1.33$) on a glass substrate. The diameter of particles is $d = 200 \, \mathrm{nm}$.
The wavelength of the laser is $\lambda = 2\pi /k = 1064 \, \mathrm{nm}$ in a vacuum.
A single focal incident laser is modeled by a Gaussian beam~\cite{Richards1959,YZhao2007,NovotnyHecht2006} with
an approximate spot size of $2\omega_0 \simeq 800 \, \mathrm{nm}$.
This is related to the maximal half-angle, $\theta_{\rm max}$, by $\omega_0 \sim 2/(k \tan \theta_{\rm max})$.
The numerical aperture is $NA = n_{\rm w} \sin( 2/(k\omega_0) )\simeq 0.996$.
The focal point is set on the substrate surface.
The incident light propagates from water to glass (see Figs.\ \ref{fig:OptBind}(a) and (b)).
The particles move and are optically manipulated on the substrate.
An effect of charge on the particles is not essential and is not considered.
The details of the methodology are summarized at the end of this Letter, after the concluding remarks.

The optical force on particle $i$ is evaluated based on the electromagnetic field on the particle surface.
The field is calculated self-consistently by the generalized Mie theory and
T-matrix method~\cite{Mackowski1994,Mackowski1996,Mackowski2011}:
\begin{equation}
\bm{E}_{\rm tot} (\bm{r}) = \bm{E}_{\rm inc} (\bm{r}) + \sum_{i=1}^{N_{\rm p}} \bm{E}_{{\rm sca},i} (\bm{r}),
\end{equation}
where $\bm{E}_{\rm inc}$ and $\bm{E}_{{\rm sca},i}$ represent the incident light and scattered light from particle $i$, respectively.
$N_{\rm p}$ is the number of particles.
The optical force $\bm{F}_i$ is calculated from the total electromagnetic field $\bm{E}_{\rm tot}$
via Maxwell's stress tensor as
\begin{equation}
\bm{F}_i = \oint_{S_i} d\Omega \cdot \left( \bar{T}_{\rm E} + \bar{T}_{\rm B} \right).
\label{eq:OptF}
\end{equation}
The integral is over the surface of particle $i$~\cite{Datsyuk2015}.
The simulation of particle dynamics follows the Langevin equation,
\begin{equation}
m \frac{d^2 \bm{r}_i}{dt^2} = -\zeta \frac{d \bm{r}_i}{dt} + \bm{F}_i + \bm{\xi}_i,
\label{eq:EOM}
\end{equation}
with $\zeta = 3\pi \eta_{\rm w} d$ being the friction coefficient.
$\bm{\xi}_i$ represents the Gaussian random force due to the Brownian motion of water molecules.
All particles are identical and their mass is $m$.
The viscosity of water is assumed to be $\eta_{\rm w} = 0.890\, \mathrm{mPa \cdot s}$ at room temperature.
The optical force on one particle accelerates the other particles due to the hydrodynamic interaction.
This is accounted for by the Ront--Prager--Yamakawa mobility tensor~\cite{Rotne1969,Yamakawa1970,Happel1983} (see method),
which gives additional velocities to particle $i$ from the external force acting on the other $j \ne i$:
\begin{equation}
\Delta \bm{v}_i = \sum_{j \ne i} \tilde{\bm{\mu}}_{ij} \bm{F}_j.
\label{eq:RPYmobility}
\end{equation}

\section{Results and discussion}
\subsection{Dynamics of gold particles}
We use the simulated results to reveal the indirect mechanism.
The multiple scattering and interference with the incident field are essential.
The binding distance approximately corresponds to the wavelength of light.
The dynamics is faster with an increase in the number of bound particles.
The observed motion is elucidated based on the ``SO coupling'' of light and the optical current~\cite{Berry2009}, as explained below.
The correlation between SAM and OAM has previously been discussed for the rotational optical manipulation by
the optical vortex with OAM~\cite{Tamura2019}.
However, it should be noted that SO coupling plays an essential role even when light with only SAM is injected
if the targeted matter system has a geometrical structure.

Figures \ref{fig:OptBind}(c)--(g) show a snapshot of the intensity of the total electric field with
$N_{\rm p} = 2 \sim 8$ particles and their trajectories when the circular polarization of the incident laser is applied.
At the center of the focal area, one particle is optically trapped directly by the incident laser.
The trapped particle causes a scattering of incident laser.
Then, due to interference, the light intensity shows a ring-shaped oscillation
(Fig.\ \ref{fig:OptBind}(c) and Fig.\ \ref{figS:intensity} in Appendix \ref{appen:field}).
This oscillation results in the binding of surrounding particles by an indirect mechanism in
the vicinity of the local maximum of the intensity at $r \simeq \lambda /n_{\rm w}$ from the center particle.
The map of the optical force on an additional small particle shown in Fig.\ \ref{figS:SFvecF}(b) in Appendix \ref{appen:ocof}
clearly indicates the position of binding due to the indirect mechanism.
How the added particles are bounded one-by-one is shown in Figs.\ \ref{fig:OptBind}(d)--(f).
For six surrounding particles, the bound position from the center is $\Delta \approx 853.4\, \mathrm{nm}$.
The intervals between the particles are approximately equivalent with $\lambda /n_{\rm w}$.
The particles show a hexagonal distribution, which indicates that our simulation well reproduces
the experimental observation in Ref.~\cite{Kudo2018}.
When $N_{\rm p} \geq 8$, we find two semi-stable distributions, as shown in Figs.\ \ref{fig:OptBind}(f) and (g).
Figure \ref{fig:OptBind}(g) shows the optical binding at the second neighboring position
(see also the binding position in Fig.\ \ref{figS:SFvecF}(d) in Appendix \ref{appen:ocof}).
This is not at the local maximum of the intensity and suggests the indirect mechanism of binding.
As a side note, the case of a linear polarized laser is discussed in Appendix \ref{appen:linear},
where the particles are aligned in a direction perpendicular to the polarization.
This result is a good demonstration that our simulation reasonably explains experimental observations~\cite{Kudo2018}.

The center particle at the focal point is strongly trapped and hardly moves,
whereas the other surrounding particles revolve by the optical force.
The lines in Figs.\ \ref{fig:OptBind}(c)--(g) indicate a slight trajectory of the revolution.
The direction of revolution accords with the optical current shown in Figs.\ \ref{figS:SFvecF}(a) and (c),
which is in contrast with the map of force on an additional small particle in Figs.\ \ref{figS:SFvecF}(b) and (d).
The speed of revolving particles increases with the number of surrounding particles.
This acceleration is attributed to an enhancement of the multiple scattering of light.
When $N_{\rm p} = 7$ in Fig.\ \ref{fig:OptBind}(e), the hexagonal distribution is approximately maintained during the revolution,
although the particles are affected by the random force.
The motions are shown in Movie of Appendix.
In Fig.\ \ref{fig:OptBind}(g), the second neighboring particle is much slower than
the first neighboring particles because of the decreased light intensity.

The optical force is evaluated by the square of the field, $|\bm{E}_{\rm inc} + \sum_i \bm{E}_{{\rm sca},i}|^2$.
The interference term is significant for the binding and revolution, whereas
$|\bm{E}_{\rm inc}|^2$ and $\sum_i |\bm{E}_{{\rm sca},i}|^2$ might result in force in the radial and out-of-plane directions.
Because the scattered field is larger than or comparable to the incident field at the binding position,
the self-assembly process is critical for indirect optical manipulation.
It has the possibility to achieve more various configurations of particles by utilizing the internal degrees of freedom of light.

\subsection{Angular momentum conversion}
For the dynamics, we analyze the SAM and OAM components of scattered light on
the upper and lower celestial hemispheres at $r = R_{\rm c} \gg d$:
\begin{eqnarray}
C_{\sigma, l} &=& C_{\sigma, l}^{\rm (u)} + C_{\sigma, l}^{\rm (l)}
\label{eq:totalC} \\
C_{\sigma, l}^{\rm (u)}
&=& \frac{1}{C^{(0)}} \int_0^{\pi /2} \sin \theta d\theta \nonumber \\
& & \hspace{-15mm} \times \left| \int_0^{2\pi} d\phi \left\{ \bm{e}_{\sigma,l} (\theta, \phi)
e^{iR_{\rm c}} \right\}^\ast
\cdot \bm{E}_{\rm tot} (R_{\rm c},\theta, \phi) \right|,
\label{eq:upperC} \\
C_{\sigma, l}^{\rm (l)}
&=& \frac{1}{C^{(0)}} \int_{\pi /2}^{\pi} \sin \theta d\theta  \nonumber \\
& & \hspace{-15mm} \times \left| \int_0^{2\pi} d\phi \left\{ \bm{e}_{\sigma,l} (\theta, \phi)
e^{iR_{\rm c}} \right\}^\ast \right.
\nonumber\\
& & \hspace{-5mm} \cdot
\left\{ \bm{E}_{\rm tot} (R_{\rm c},\theta, \phi)  -\bm{E}_{\rm inc} (R_{\rm c},\theta, \phi) \right\} \Big|.
\label{eq:lowerC}
\end{eqnarray}
Here, $\bm{e}_{\sigma,l} (\theta, \phi)$ corresponds to the mode of the electric field with
spin $\sigma$ and vortex $l$ with respect to the $z$-axis (see definition in Appendix \ref{appen:esl}).
$C^{(0)}$ is a normalization factor to satisfy $\sum_{\sigma ,l} {C_{\sigma ,l}}^2 =1$.
Note that the subtraction in Eq.\ (\ref{eq:lowerC}) is to consider only the emission.

In the case of a plane wave, the light has no OAM ($l = 0$)~\cite{YZhao2007}.
On the contrary, a focal laser with right circular polarization has a slight $(\sigma ,l) = (-1,+2)$ component
in addition to the majority $(+1,0)$ component.
This is due to ``scattering'' by the lens.
The incident laser consists of $C_{+1,0} \approx 0.9750$ and $C_{-1,+2} \approx 0.2222$ when $2\omega_0 = 800\, \mathrm{nm}$.
Note that the sum of their squares is unity.

\begin{figure}[t]
\includegraphics[width=1.0\linewidth]{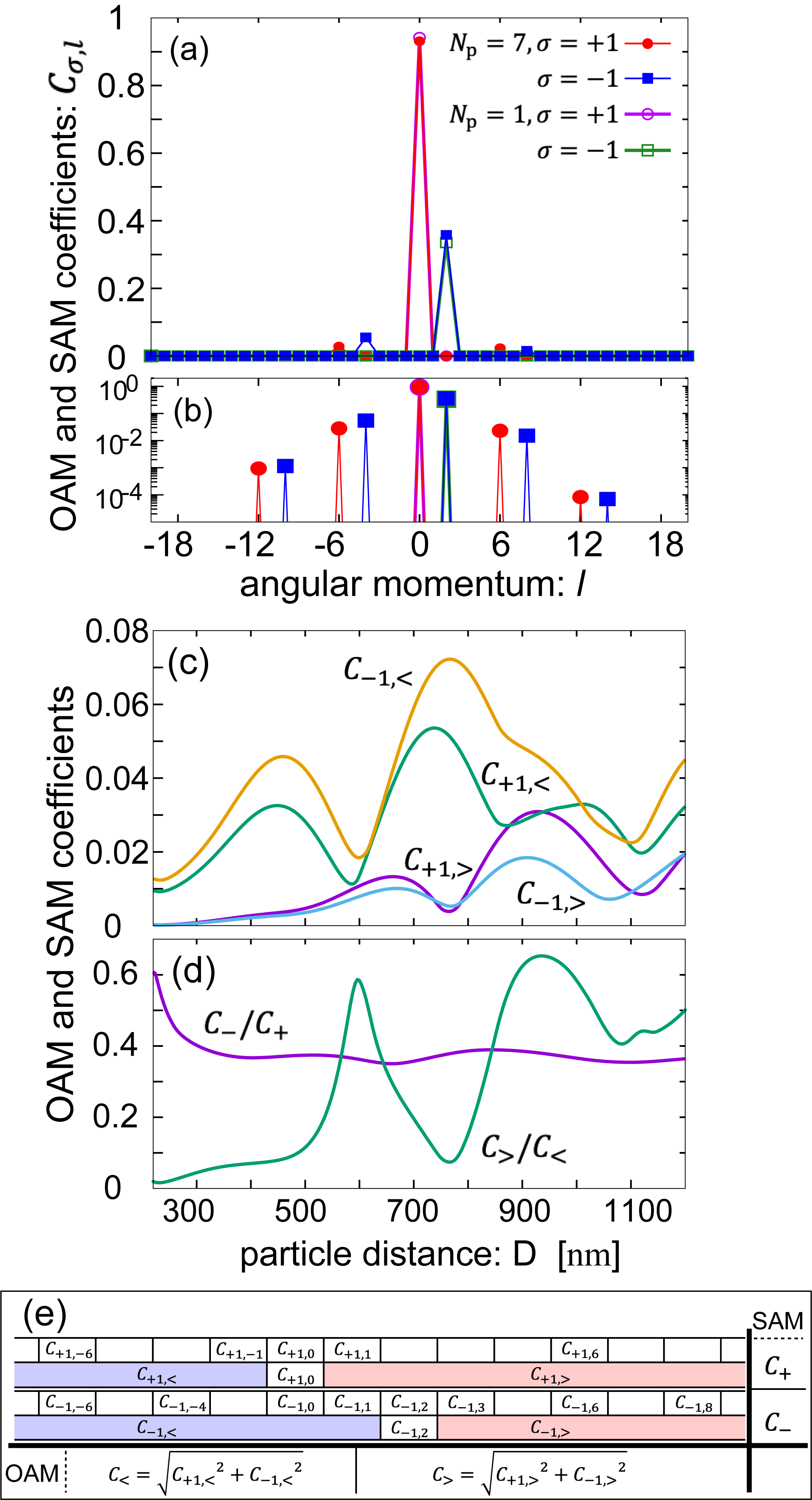}
\caption{
(a) Coefficients of SAM and OAM components of scattered field with $N_{\rm p} = 1,7$
when $\sigma = +1$ circular polarized light is tightly focused and applied.
$N_{\rm p} =7$ and the other parameters correspond to Fig.\ \ref{fig:OptBind}(e).
(b) Log scale plot of (a).
(c,d) Particle distance dependence of the coefficients when $N_{\rm p} =7$.
(e) Summary of the coefficients for the SAM and OAM defined in Eqs.\ (\ref{eq:cpm}) and (\ref{eq:cgl})
and the sentence that follows them.
}
\label{fig:OAMSAM}
\end{figure}

Scattering by a particle or particle assembly produces a further $C_{-1,+2}$ component.
Figures \ref{fig:OAMSAM}(a) and (b) exhibit the coefficient $C_{\sigma ,l}$ of light scattered when $N_{\rm p} =1$ and $7$.
For a single particle, the scattered light shows $C_{-1,+2} > 0.3$, and the others of $\sigma = -1$ are zero.
This represents a conversion of the SAM to the OAM under the conservation of total angular momentum (TAM), $j = l + \sigma$.
This is regarded as SO coupling for the light.
However, it is difficult to determine the orbital motion due to a strong trapping.

In the situation of Figs.\ \ref{fig:OptBind}(c)--(g), the rotational symmetry is broken and the TAM of light is not conserved.
In Fig.\ \ref{fig:OptBind}(e), the system has a six-hold rotational symmetry and
the scattered light consists of $l= \pm 6 m$ with $\sigma = +1$ (where $m$ is an integer).
A spin-flip also occurs, and $C_{-1,2 \pm 6m}$ is generated, where the TAM is distributed from $j=1$ to only $1 \pm 6 m$.
Here, note that the distribution of TAM (mainly OAM) is imbalanced and, the coefficient of $j=1-6$ is larger than that of $j=1+6$.
This is consistent with the rotation of assembly if the particles and photons follow Newton's third law of motion.

Here, we claim that the key elements of the particle dynamics are the spin-flip by
the ``SO coupling'' and the imbalance of OAM generation.
Let us discuss the SAM and OAM components in the parameter space of the particle distance and the complex refractive index.
We introduce
\begin{eqnarray}
C_{\pm}  &=& \left[ {\textstyle \sum_l} {C_{\sigma = \pm 1, l}}^2 \right]^{\frac{1}{2}},
\label{eq:cpm} \\
C_{> (<)} &=& \left[ {C_{+1, > (<)}}^2 + {C_{-1, > (<)}}^2 \right]^{\frac{1}{2}}
\label{eq:cgl}
\end{eqnarray}
with ${C_{+1, > (<)}} = [ \sum_{l>0 (<0)} {C_{+1,l}}^2 ]^{1/2}$ and
${C_{-1, > (<)}} = [ \sum_{l>2 (<2)} {C_{-1,l}}^2 ]^{1/2}$.
Note that $C_{\pm}$ and $C_{> (<)}$ are indicators of the extent of the spin-flip and imbalance of
OAM generation, respectively, which are summarized in Fig.\ \ref{fig:OAMSAM}(e).

First, we examine the case that the particle distance $\Delta$ changes virtually, as shown in Figs.\ \ref{fig:OAMSAM}(c) and (d).
The oscillations of the coefficients are seen in these figures.
Here, we can see why the rotation is strongly driven in the considered system.
Figure\ \ref{fig:OAMSAM}(c) shows that the spin-flip coefficients generating negative $l$, namely $C_{\pm 1,<}$,
are enhanced around $\Delta$ corresponding to the light wavelength in water and its half, 
whereas the maximum positions of the coefficients generating positive $l$,
namely $C_{\pm 1,>}$ are shifted, i.e., oscillate in a different phase, and $C_{\pm 1,>}$
becomes minimum around $\Delta \approx$ wavelength.
This is why the imbalance of OAM becomes maximum (namely, $C_{>}/C_{<}$ becomes minimum) there.
At the binding position of our calculation, the OAM is generated with sufficient imbalance to
drive the revolution of surrounding particles in a particular direction. 
The oscillation period suggests that the interference of multiple scattering of light by the particle's configuration
determines the generation of OAM.

Figure\ \ref{fig:OAMSAM}(d) shows that $C_{-}/C_{+}$ is insensitive to $\Delta$,
except in the region of small $\Delta$, and that $C_{>} < C_{<}$ is maintained.
The latter observation agrees with the absence of negative torque for a tightly focused laser,
which is in contrast with the case of wide-area irradiation~\cite{FHan2018}.
In the case of the wide-area irradiation, the interference between the incident and first-order scattering fields from
the respective particles is significant rather than the multiple scattering in the present case;
the maps of optical current and force are shown in Fig.\ \ref{figS:SFvecW}.
The rotation behavior arising from the interference of scattered light can also be discussed in terms of the optical current.
The relevant figures and discussion are provided in Appendix \ref{appen:ocof}.

\begin{figure}[t]
\includegraphics[width=1.0\linewidth]{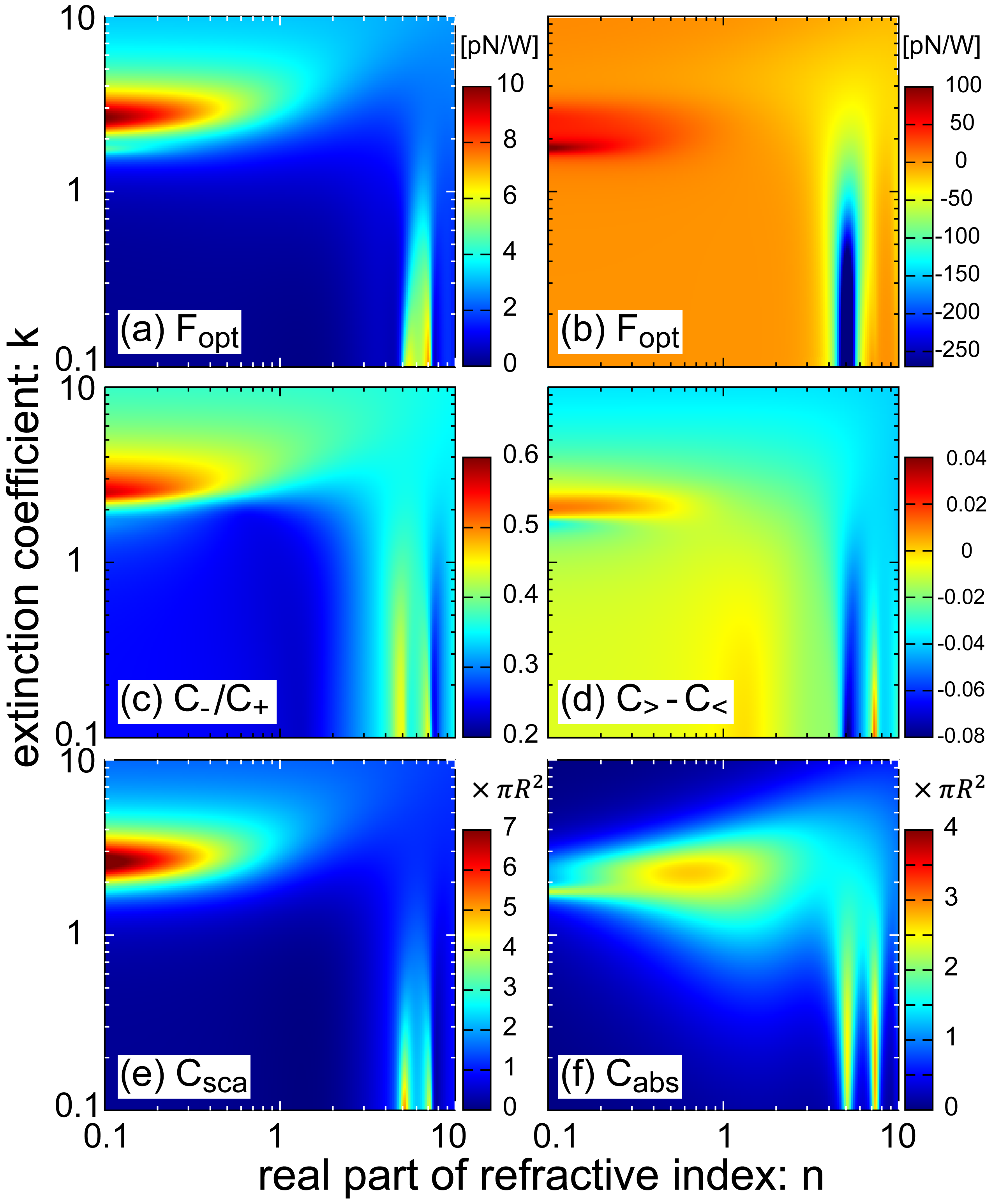}
\caption{
(a,b) Profile of in-plane optical force $(F_x,F_y)$ on one of the surrounding particles in the plane of complex refractive index
$\tilde{n} = n + i\kappa$ when $N_{\rm p} =7$ and tightly focused (a) and not focused (b) lasers are applied.
The beam waists are $2\omega_0 = 0.8\, \mathrm{\mu m}$ and $12.5\, \mathrm{\mu m}$, respectively.
The latter situation is a conventional setup for optical binding.
The diameter of particles is $d=200\, \mathrm{nm}$.
One particle is trapped at $r=0$, and the others are at $r=\Delta = 853.4\, \mathrm{nm}$ (see Fig.\ \ref{fig:OptBind}(e)).
(c,d) Spin-flip ratio due to the ``SO coupling'' (c) and imbalance of OAM generation (d) evaluated from the coefficients of SAM and OAM.
(e,f) Single particle cross-section of the scattering $C_{\rm sca}$ (e) and absorption $C_{\rm abs}$ (f).
The plot is normalized by $\pi R^2$, where $R$ is the radius.
}
\label{fig:NKplane}
\end{figure}

As shown in Fig.\ \ref{fig:OAMSAM} (d), $C_{-}/C_{+}$ is insensitive to the particle distance.
However, it remarkably depends on the particle material, specifically its complex refractive index, as discussed below. 
Here, we discuss the spin-flip and OAM imbalance ratios by examining the indirect optical force,
scattering, and absorption cross-sections in a plane of the complex refractive index.
Figure \ref{fig:NKplane}(a) exhibits the indirect optical force driving the surrounding particles in
the situation of Fig.\ \ref{fig:OptBind}(e).
Positive force corresponds to the counterclockwise direction in Fig.\ \ref{fig:OptBind}(e).
The optical force is enlarged when the refractive index is almost purely imaginary (perfect conductor),
and the imaginary part is $\kappa \approx 2.67$.
The force is also enlarged around $n \approx 6.5, \kappa \ll 1$.

For a tightly focused laser, the indirect optical force is only positive in the $n$-$\kappa$ plane,
whereas both positive and negative force is found when the light is applied widely [Fig.\ \ref{fig:NKplane}(b)].
This is a clear difference between indirect and direct optical manipulation.
The negative force is enlarged at $n \approx 5.1, \kappa \ll 1$.
The presence of negative torque in direct optical binding with wide-area irradiation is consistent with Ref.~\cite{FHan2018}.

The ``SO coupling'' is also examined in the $n$-$\kappa$ plane for the seven particles in Fig.\ \ref{fig:NKplane}(c).
The profile of spin-flip ratio $C_- /C_+$ shows similar behavior to the indirect optical force,
where a region of spin-flip enhancement correlates with a large indirect force region.
Meanwhile, the difference of generated OAM, $C_> - C_<$, in Fig.\ \ref{fig:NKplane}(d) shows that the sign changes and
there are positive and negative peaks at $n \ll 1, \kappa \approx 6.5$ and $n \approx 8.6, \kappa \ll 1$, respectively.
This result indicates that negative torque can appear if the material is different, even in the case of focused irradiation.

Figures \ref{fig:NKplane}(e) and (f) exhibit the scattering and absorption cross-sections of a
single particle in water, respectively, evaluated by the Mie coefficients $a_n$ and $b_n$~\cite{BohrenHuffman1998} as follows:
\begin{eqnarray}
C_{\rm sca} &=& \frac{2\pi}{k^2} \sum_{n=1}^{\infty} (2n+1) (|a_n|^2 + |b_n|^2),
\label{eq:Csca} \\
C_{\rm ext} &=& \frac{2\pi}{k^2} \sum_{n=1}^{\infty} (2n+1) {\rm Re} ( a_n + b_n ),
\label{eq:Cext} \\
C_{\rm abs} &=& C_{\rm ext} - C_{\rm sca}.
\label{eq:Cabs}
\end{eqnarray}
The profile of $C_{\rm sca}$, especially in the enhancement parameter regions, shows good agreement with
the indirect optical force in Fig.\ \ref{fig:NKplane}(a).
Meanwhile, $C_{\rm abs}$ is not enlarged at $n \ll 1$.
Therefore, by comparing Fig.\ \ref{fig:NKplane}(a) with (c) and (e), we can see that the enhancements of scattering
and ``SO coupling'' are dominant factors in the indirect optical manipulation.

The above discussion gives a guideline for indirect and wide-area optical manipulation by multiple scattering,
such that one should prepare particles and conditions with large scattering cross-section rather than absorption.
To transfer the momentum or angular momentum of photons to the particles directly, both strong scattering and absorption are suitable.
However, if particles show a strong absorption, the light is extinct quickly and multiple scattering is suppressed.
For the case of $d=200\, \mathrm{nm}$, $C_{\rm sca} \gg C_{\rm abs} \approx 0$ is a better condition for indirect optical manipulation.
The particle size is also an important parameter.
In Appendix \ref{appen:size}, we investigate the indirect force and cross-sections for various diameters (see Fig.\ \ref{figS:FCsCa}).
We find that the criterion to obtain large indirect optical force is adaptable for $d \gtrsim 150\, \mathrm{nm}$.

\section{Conclusions}
In conclusion, we conducted numerical simulations of nanoparticles optically trapped and bound by a single focused laser.
The simulation revealed that the scattered light from the strongly trapped center particle causes
binding and indirect optical force on the surrounding particles.
Due to the multiple scattering between all particles, the ordering and optical manipulation of particles can be achieved
beyond the focal irradiation area.
Under circularly polarized laser irradiation, a hexagonal ordering of particles with wavelength distance is formed,
and the surrounding particles revolve.
The simulated results qualitatively agree well with recent experimental observations~\cite{Kudo2018}.

Based on the analysis of SAM and OAM, we revealed the mechanisms to enlarge the revolution in the indirect optical manipulation:
$C_{\rm sca} \gg C_{\rm abs}$ and large ``SO coupling'' of light, which must be related with each other.
How a strong SO coupling contributes to the revolution is also elucidated schematically in Fig.\ \ref{figS:OAM02} in Appendix \ref{appen:SO}.
This is determined by the particle properties , i.e., complex refractive index, diameter, shape, etc.
Their engineering may be possible by using core-shell structures or dye-doped polymers.
The imbalance of OAM generation is another important factor for both indirect and direct optical manipulation,
which is largely affected by the particle placement.

As we discuss in Appendix \ref{appen:ocof}, the optical binding and dynamics of particles are
described well by the optical force and current, respectively.
The structure of optical current is induced by interference between the scattering and incident fields.
Thus, strong scattering enlarges the dynamics.

The present results indicate the possibility that a scanning single beam can create an ordered pattern with
local structures over a wide area by controlling the polarization.
If we use several beams simultaneously with different phases,
the degrees of freedom for designing the pattered structures will be greatly enhanced.
We suggest that such large degrees of freedom would open the possibility of new optical manipulation,
such as a phase transition of the order of optical binding, as will be elaborated on in our next publication.

\section{Method}
The remainder of this Letter details our methodology.
We simulated the dynamics of nanoscale spherical gold particles in a water solvent on a glass substrate.
A single focal incident laser with counterclockwise circular polarization was considered.
The electric field was modeled as a Gaussian beam~\cite{Richards1959,YZhao2007,NovotnyHecht2006}:
\begin{equation}
\bm{E}_{\rm inc} (r,\varphi ,z)
= -\frac{ikf}{2} \sqrt{\frac{n_1}{n_2}} E_0 e^{-ikf}
\left[ \begin{matrix}
  I_{00} + I_{02} e^{i2\varphi}  \\
i (I_{00} - I_{02} e^{i2\varphi}) \\
-2i I_{01} e^{i\varphi}
\end{matrix} \right]
\end{equation}
with
\begin{eqnarray}
I_{00} &=& \int_0^{\theta_{\rm max}} d\theta
f_w (\theta) \sqrt{\cos \theta} \sin \theta (1+\cos \theta) \nonumber\\
&& \times J_0 (kr \sin \theta) e^{ikz \cos \theta}, \\
I_{01} &=& \int_0^{\theta_{\rm max}} d\theta
f_w (\theta) \sqrt{\cos \theta} \sin^2 \theta \nonumber\\
&& \times J_1 (kr \sin \theta) e^{ikz \cos \theta}, \\
I_{02} &=& \int_0^{\theta_{\rm max}} d\theta
f_w (\theta) \sqrt{\cos \theta} \sin \theta (1-\cos \theta) \nonumber\\
&& \times J_2 (kr \sin \theta) e^{ikz \cos \theta},
\end{eqnarray}
where $f_w (\theta) = \exp ( - (\sin \theta /\sin \theta_{\rm max})^2)$.
$J_n (x)$ is the $n$-the order Bessel function.
The parameters $f$, $\theta_{\rm max}$, and $k$ represent the focal distance,
maximal half-angle of light cone, and wavenumber in a vacuum, respectively.

The incident field is expanded by the vector spherical harmonics (VSH) functions, $\bm{M}_{nmp,i}^{(1)}$ and $\bm{N}_{nmp,i}^{(1)}$, as
\begin{eqnarray}
\bm{E}_{\rm inc} (\bm{r}) &=& \sum_{n=1}^{n_{\rm max}} \sum_{m=-n}^{+n} \sum_{p = {\rm e,m}}
\left[ u_{nmp,i} \bm{M}_{nmp,i}^{(1)} (\bm{r}) \right.\nonumber\\
&+& \left. v_{nmp,i} \bm{N}_{nmp,i}^{(1)} (\bm{r}) \right].
\end{eqnarray}
Here, the superscript $(1)$ in $\bm{M}_{nmp,i}^{(1)}$ and $\bm{N}_{nmp,i}^{(1)}$ represents
the spherical Bessel function in the radial part to describe the incident field.
The subscript $i$ denotes that these VSH functions are located at $\bm{r}_i$, i.e., the position of particle $i$.
The index $p = {\rm e,m}$ corresponds to the TE and TM modes.
$u_{nmp,i}$ and $v_{nmp,i}$ are the expansion coefficients, as determined by localized approximation~\cite{Mackowski1994}.
The expansion of $\bm{E}_{\rm inc}$ by $\bm{M}_{nmp,i}^{(1)}$ and $\bm{N}_{nmp,i}^{(1)}$ is applied to its scattering by particle $i$.

When the particles are isolated from each other, the scattered field is approximately given by
\begin{eqnarray}
\bm{E}_{{\rm sca},i}^{(0)} (\bm{r}) &=& \sum_{n=1}^{n_{\rm max}} \sum_{m=-n}^{+n}
\left[ a_n u_{nmp,i} \bm{M}_{nmp,i}^{(3)} (\bm{r}) \right. \nonumber\\
&+& \left. b_n v_{nmp,i} \bm{N}_{nmp,i}^{(3)} (\bm{r}) \right]
\end{eqnarray}
with the Mie coefficients $a_n$ and $b_n$ being independent of $i$.
Note that the radial part of $\bm{M}_{nmp,i}^{(3)}$ and $\bm{N}_{nmp,i}^{(3)}$ is given by
the spherical Hankel function of the first kind to describe the outward spherical wave.
By introducing vectors $\vec{c}_{{\rm inc},i} = (\{ u_{nmp,i}, v_{nmp,i} \})^{\rm t}$ and
$\vec{c}_{{\rm sca},i}^{\ (0)} = (\{ a_n u_{nmp,i}, b_n v_{nmp,i} \})^{\rm t}$ for the incident and scattered fields, respectively,
the Mie coefficients give the T-matrix for a single particle, $\vec{c}_{{\rm sca},i}^{\ (0)} = \hat{t}_i \vec{c}_{{\rm inc},i}$.

The multiple scatterings between the particles combine the T-matrices $\hat{t}_i$ and
result in generalized T-matrix elements $\hat{T}_{ij}$~\cite{Mackowski1994,Mackowski1996,Mackowski2011}.
An extension of the vector, $\vec{C}_{\rm inc} = (\vec{c}_{{\rm inc},1}, \vec{c}_{{\rm inc},2}, \cdots)^{\rm t}$,
shows a simple formulation of the multiple scattering:
\begin{equation}
\vec{C}_{\rm sca} = \left( \begin{matrix}
\vec{c}_{{\rm sca},1} \\
\vec{c}_{{\rm sca},2} \\
\vdots
\end{matrix} \right)
= \left( \begin{matrix}
\hat{T}_{11} & \hat{T}_{12} & \\
\hat{T}_{21} & \hat{T}_{22} & \\
             &              & \ddots
\end{matrix} \right) \vec{C}_{\rm inc}.
\end{equation}
The evaluated coefficients $\vec{c}_{{\rm sca},i} = (\{ A_{nmp,i}, B_{nmp,i} \})^{\rm t}$ give a full scattered field
\begin{eqnarray}
\bm{E}_{{\rm sca},i} (\bm{r}) &=& \sum_{n=1}^{n_{\rm max}} \sum_{m=-n}^{+n}
\left[ A_{nmp,i} \bm{M}_{nmp,i}^{(3)} (\bm{r}) \right. \nonumber\\
&+& \left. B_{nmp,i} \bm{N}_{nmp,i}^{(3)} (\bm{r}) \right].
\end{eqnarray}

\section*{ACKNOWLEDGMENT}

The authors thank Prof.\ H.\ Masuhara, Dr.\ T.\ Kudo, and Z.-H.\ Huang for their fruitful discussions on their experimental results.
The authors acknowledge for all member of Collective Optofluidic Dynamics of Nanoparticles meeting organized by Prof.\ Masuhara.
This work was supported by JSPS KAKENHI Grant Number 16K21732 in Scientific Research on Innovative Areas ``Nano-Material Optical-Manipulation''.
T.Y.\ was supported by JSPS KAKENHI Grant Number 18K13484 and H.I.\ was  supported by JSPS KAKENHI Grant Number 16H06504 and 18H01151.

\appendix
\section{Profile of incident and total fields}
\label{appen:field}
\begin{figure}[t]
\includegraphics[width=0.8\linewidth]{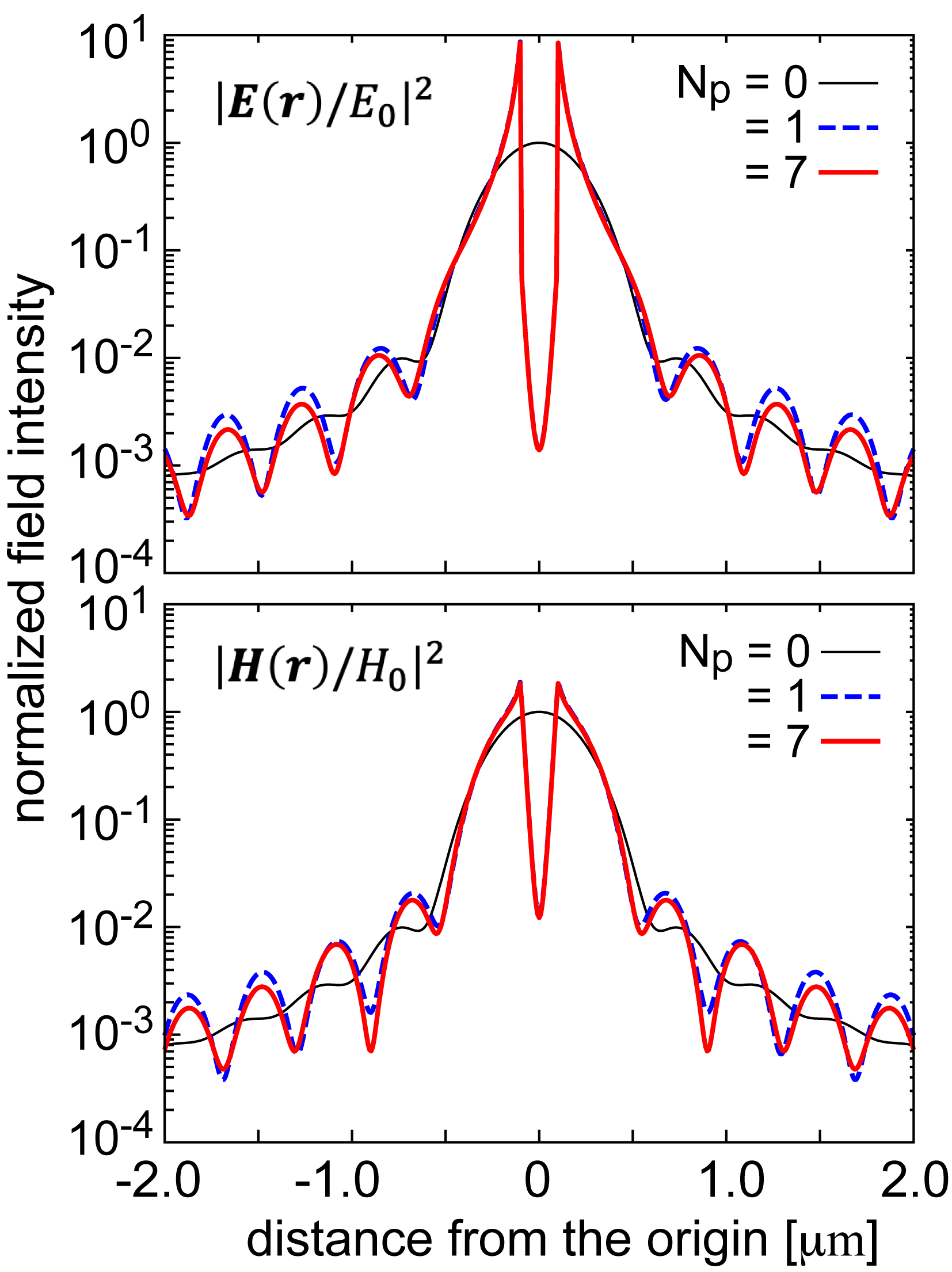}
\caption{
Profile of the intensities of total electric and magnetic fields, $|\bm{E}_{\rm tot}|^2$ and $|\bm{H}_{\rm tot}|^2$,
for a circular polarization when $N_{\rm p} = 0,1$, and $7$.
The focal point of the incident laser is located at the origin.
The intensities are normalized by the incident field at the focal point,
$|\bm{E}_{\rm inc} (r=0)|^2 = {E_0}^2$ and $|\bm{H}_{\rm inc} (r=0)|^2 = {H_0}^2$.
The particles are gold with $d=200\, \mathrm{nm}$ diameter.
}
\label{figS:intensity}
\end{figure}

\addcontentsline{toc}{section}{Profile of incident and total fields \dotfill}
In our discussion on indirect optical manipulation under tightly focused laser irradiation,
the scattered field by a central trapped particle is larger than or comparable with the incident field at
the binding position of surrounding particles.
Therefore, the multiple scattering of light between particles is significant.
To show the field profile explicitly in our setup with $NA \simeq 0.996$, we plot the intensities of
total electric fields in Fig.\ \ref{figS:intensity} when $N_{\rm p} = 0$ (incident field), $1$, and $7$.
The incident field is a Gaussian beam applying localized approximation with the Gaussian beam constant $C_{\rm GB}$ being $0.51$.
Although this value is beyond an appropriate range of approximation with $C_{\rm GB} = 1/(k \omega_0)$ to
estimate the beam waist $\omega_0$, it is chosen to produce $\omega_0 \approx 800\, \mathrm{nm}$;
see the plot of $N_{\rm p} = 0$ in Fig.\ \ref{figS:intensity}.

When the light is applied to a single particle, the intensity oscillates as a function of
the distance $r$ from the focal center due to interference between the incident and scattered fields.
At $r \simeq 850\, \mathrm{nm}$, the intensity indicates the first maximum,
where the gradient force is zero and the particle could be bound optically.
At the local maximum, the total field intensity is larger than the incident one.
In the case of $N_{\rm p} = 7$, the profile plot of field intensity follows a bisector between
two of the surrounding particles (for the case of Fig.\ \ref{fig:OptBind}(e), along the $y$ axis).
The oscillations of intensity along the radial direction for $N_{\rm p} = 1$ and $7$ are almost equivalent.

We also plot the intensity of the magnetic field.
The magnetic field intensity is also larger than or comparable with that of incident light.
Therefore, a scattering-induced component of the Poynting vector describing the optical current is
comparable with the component due to the incident field.

\section{Optical current vs. optical force}
\label{appen:ocof}
\begin{figure*}[t]
\includegraphics[width=0.8\linewidth]{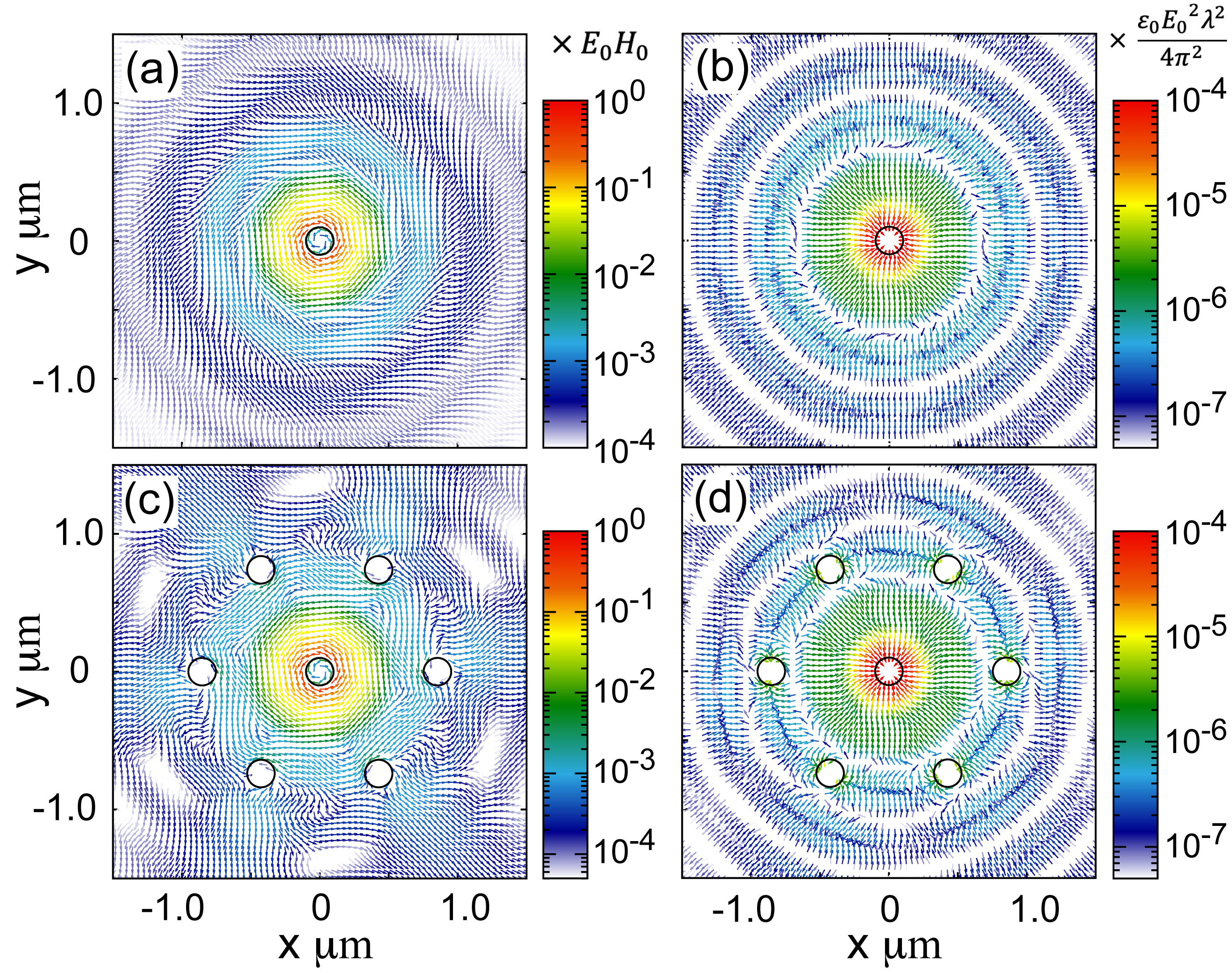}
\caption{
Poynting vectors indicating the optical current (left) and optical force on a small additional particle (right panels)
in the $z=0$ plane when $N_{\rm p} = 1$ (a,b) and $7$ (c,d) under tightly focused laser irradiation.
The incident laser has $\sigma = +1$ circular polarization. The beam waist is $0.8\, \mathrm{\mu m}$.
The other parameters of the laser and particles are the same as those in Fig.\ \ref{fig:OptBind}(e) in the main text.
In the right panels, the additional particle is $d=10 \, \mathrm{nm}$ to avoid multiple scattering by the additional particle.
The color scale displays the absolute value of vectors, $\sqrt{F_x^2 + F_y^2}$.
The black circle indicates Au particles with $d=200 \, \mathrm{nm}$.
}
\label{figS:SFvecF}
\end{figure*}

\addcontentsline{toc}{section}{Optical current vs. optical force\dotfill}
In the main text and Fig.\ \ref{fig:OAMSAM}, we examine the analysis of scattered light in terms of the SAM and OSM to
discuss the mechanism of revolution by the indirect optical force.
In our simulation with gold particles and a tightly focused laser, the imbalance of generated OAM
is consistent with the revolution direction.
To consider a mechanism to determine the revolution direction, as shown in Fig.\ \ref{figS:SFvecF},
we examine maps of the optical current described by the Poynting vector $\bm{S} = \bm{E}^* \times \bm{H} /2$
and the optical force acting on an additional small particle.
Note that the optical force acts on all the present particles.
To consider an `` optical force at an arbitrary position'', we introduce a particle,
which is $d=10 \, \mathrm{nm}$ to avoid additional light scattering.
The optical current describes the contribution of the scattering force, whereas
the optical force consists of both the scattering and gradient forces.

The optical current of the incident laser is almost along the incident direction, and
the in-plane components are negligible even for a tightly focused laser (not shown).
With the scattered field, the optical current exhibits the spatial structures.
When one particle is trapped at the focal point, the optical current of the total field shows a vortical structure
according to the incident $\sigma =+1$ polarization in Fig.\ \ref{figS:SFvecF}(a),
whereas the optical force is in the radial direction in Fig.\ \ref{figS:SFvecF}(b).
This is due to much stronger gradient force than the scattering force.
The optical force indicates the binding position clearly, while the optical current does not.
Therefore, for the binding, one should discuss the optical force, whereas
for the dynamics or indirect manipulation, the optical current gives reasonable information.
The optical current with seven particles in Fig.\ \ref{figS:SFvecF}(c) shows only slight difference from
that with one particle in (a) at a point of the vortical flow to explain the revolution.
The qualitative difference is an enhancement of the optical currents by multiple scattering in
the vicinity of surrounding particles.
This result qualitatively explains the direction of revolution in Fig.\ \ref{fig:OptBind}(e) in the main text,
whereas the force in Fig.\ \ref{figS:SFvecF}(d) is not useful to discuss the dynamics.

\begin{figure*}[t]
\includegraphics[width=0.8\linewidth]{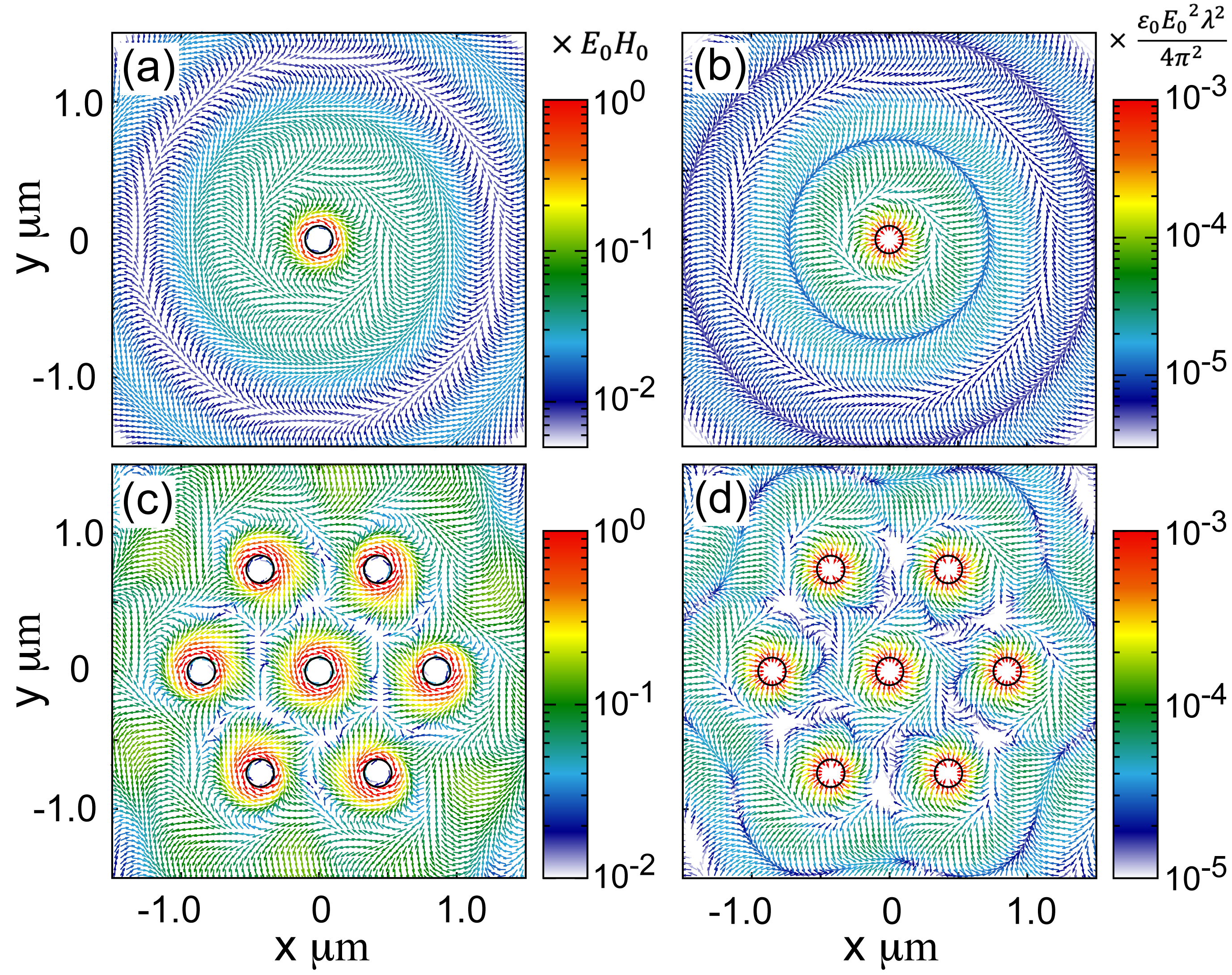}
\caption{
Poynting vectors indicating the optical current (left) and optical force on a small additional particle (right panels)
in the $z=0$ plane when $N_{\rm p} = 1$ (a,b) and $7$ (c,d) under wide-area irradiation.
The incident laser has $\sigma = +1$ circular polarization. The beam waist is $12.5\, \mathrm{\mu m}$.
The other parameters of the laser and particles are the same as those in Fig.\ \ref{fig:OptBind}(e) in the main text.
The upper four panels show the case of tightly focused laser irradiation. The lower four panels are the non-focused case.
In the right panels, the additional particle is $d=10 \, \mathrm{nm}$ to avoid multiple scattering by the additional particle.
The color scale displays the absolute value of vectors, $\sqrt{F_x^2 + F_y^2}$.
The black circle indicates Au particles with $d=200 \, \mathrm{nm}$.
}
\label{figS:SFvecW}
\end{figure*}

Figures \ref{figS:SFvecW}(a)--(d) demonstrate widely focused irradiation.
When $N_{\rm p}=1$ in Fig.\ \ref{figS:SFvecW}(a), the interference between the incident and scattered fields indicates
the changes of vortex direction of the optical current with the radial distance $r$ from the focal point.
This is also found in the optical force in Fig.\ \ref{figS:SFvecW}(b).
Such behaviors qualitatively agree with the observation of negative torque reported in Ref. ~\cite{FHan2018}.
When $N_{\rm p} =7$, the optical current implies a rotation of individual particles in Fig.\ \ref{figS:SFvecW}(c),
whereas the dynamics of whole assembly is not readable.
The optical force in Fig.\ \ref{figS:SFvecW}(d) shows the next binding positions,
which are slightly different from the second stable position in Fig.\ \ref{figS:SFvecW}(b).
Neither Figs.\ \ref{figS:SFvecW}(c) nor (d) exhibit binding or revolution, unlike the tightly focused case.
For the direct optical manipulation with a wide irradiation, one should first examine the optical current and force by
the scattering from a single particle rather than multiple scattering.

The optical current suggests the absence and presence of negative torque by tight and wide focusing, respectively.
Thus, negative torque would be obtained by properly tuning the particle distance by a given charge or other techniques.

\section{Linear polarization}
\label{appen:linear}
\addcontentsline{toc}{section}{Linear polarization\dotfill}
In contrast to circular polarization, a linear polarized laser does not have SAM.
Then, indirect optical manipulation by linear polarization must show a qualitative difference from
the circular polarized laser discussed in the main text.
Here, we consider the indirect optical binding by a focused linear polarized laser and 
demonstrate a stable alignment of multiple particles, which agrees with the experiment by Kudo {\it et al.}~\cite{Kudo2018}.

\begin{figure*}[t]
\includegraphics[width=0.7\linewidth]{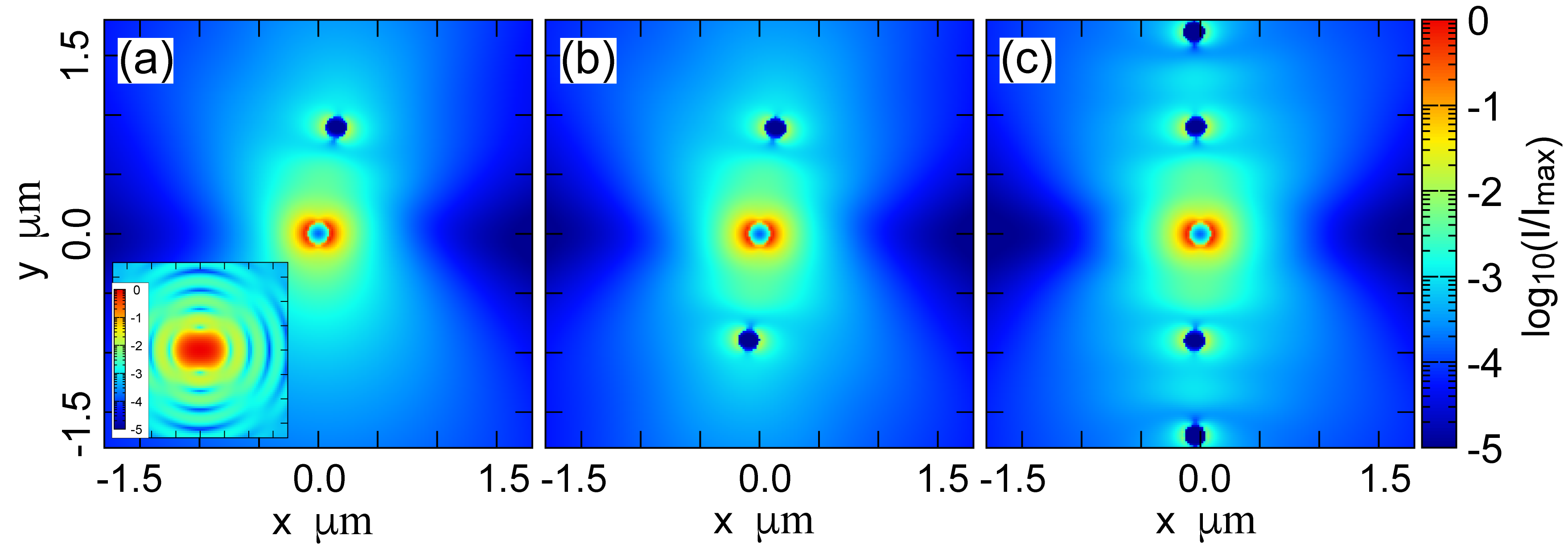}
\caption{
Intensity of total electric field with $N_{\rm p} = 2$ (a), $3$ (b), and $5$ (c) on the $z=0$ plane
when an $x$-linearly polarized and focused single laser is applied.
The focal point of the incident laser is located at the origin. The particles are stably trapped.
The inset of panel (a) shows the incident field.
}
\label{figS:linear}
\end{figure*}

The parameters of the laser are the same as those for circular polarization.
Figure \ref{figS:linear} shows the stable position of bound particles when $N_{\rm p}$ increases one by one.
The polarization direction is the $x$-direction.
In the case of linear polarization, the bound particles are fixed in a finite period and change their positions by the random force.
Thus, Fig.\ \ref{figS:linear} shows an example of possible configurations.
When $N_{\rm p} = 2$, the ``second'' particle is trapped at a slightly shifted position from the perpendicular direction
(on the $y$-axis), which may be attributed to the shape of incident laser shown in the inset of Fig.\ \ref{figS:linear}(a).
We can also find other stable trapping positions.
The ``third'' particle is trapped at another position.
When the number of particles increases, their shifts from the $y$-axis are reduced and they tend to be aligned on a line.
Such configurations were found in an experiment~\cite{Kudo2018}.

For the circular polarization, we investigate the analysis of scattered fields in terms of the SAM and OAM
and reveal that the ``SO coupling'' of light and imbalance of the generated OAM are
significant for the revolution of the surrounding particles.
Meanwhile, for linear polarization, the indirect binding is static.
Although the scattering of linear polarized light by the bound particles also causes OAM, it is balanced,
which follows the stable optical binding shown in Fig.\ \ref{figS:linear}.

\section{Definition of $\bm{e}_{\sigma,l} (\theta, \phi)$ for scattered field}
\label{appen:esl}
\addcontentsline{toc}{section}{Definition of $\bm{e}_{\sigma,l} (\theta, \phi)$ for scattered field \dotfill}
We analyze the scattered light from all particles in terms of the SAM and OAM.
The scattered light is a spherical wave at a sufficiently far position.
Therefore, to define the axis of angular momenta, the scattered field must be converted to a plane wave.
We consider a fictitious lens for this conversion.
Through the lens, the field at ($r =R_{\rm c}, \theta, \phi$) is rewritten as
\begin{equation}
\bm{E}_{\rm tot} \to  \hat{R}_y^{-1} \hat{R}_z^{-1} \bm{E}_{\rm tot}
\end{equation}
with
\begin{eqnarray}
\hat{R}_z (\theta, \phi) &=&
\left( \begin{matrix}
\cos \phi & -\sin \phi & 0 \\
\sin \phi & \cos \phi & 0 \\
0 & 0 & 1
\end{matrix} \right), \\
\hat{R}_y (\theta, \phi) &=&
\left( \begin{matrix}
\cos \phi & 0 & \sin \phi  \\
0 & 1 & 0 \\
-\sin \phi & 0 & \cos \phi
\end{matrix} \right)
\end{eqnarray}
being the rotation matrices with respect to the $z$ and $y$-axes.
The converted field is projected onto $\bm{e}_{\sigma = \pm 1} = (1,\pm i,0)^{\rm t} / \sqrt{2}$ for
$\sigma$ SAM with $l$ OAM, $\bm{e}_{\sigma} e^{il\phi}$.
Thus, the unit vector $\bm{e}_{\sigma,l} (\theta, \phi)$ in Eqs.\ (6) and (7) is given as
\begin{equation}
\bm{e}_{\sigma,l} (\theta, \phi) = \hat{R}_z \hat{R}_y \bm{e}_{\sigma} e^{il\phi}.
\end{equation}

\section{Particle size dependence of indirect optical force and cross-sections}
\label{appen:size}
\addcontentsline{toc}{section}{Particle size dependence of indirect optical force and cross-sections \dotfill}
In the main text, we discussed the correlation of the indirect optical force and cross-sections in the $n$-$\kappa$ plane
when the particle diameter is fixed at $d = 200\, \mathrm{nm}$ in Fig.\ \ref{fig:NKplane}.
Then, we noted that for the indirect mechanism, the scattering rather than the absorption is essential.
Here, we examine this criterion when the particle size is changed.

\begin{figure*}[t]
\includegraphics[width=1.0\linewidth]{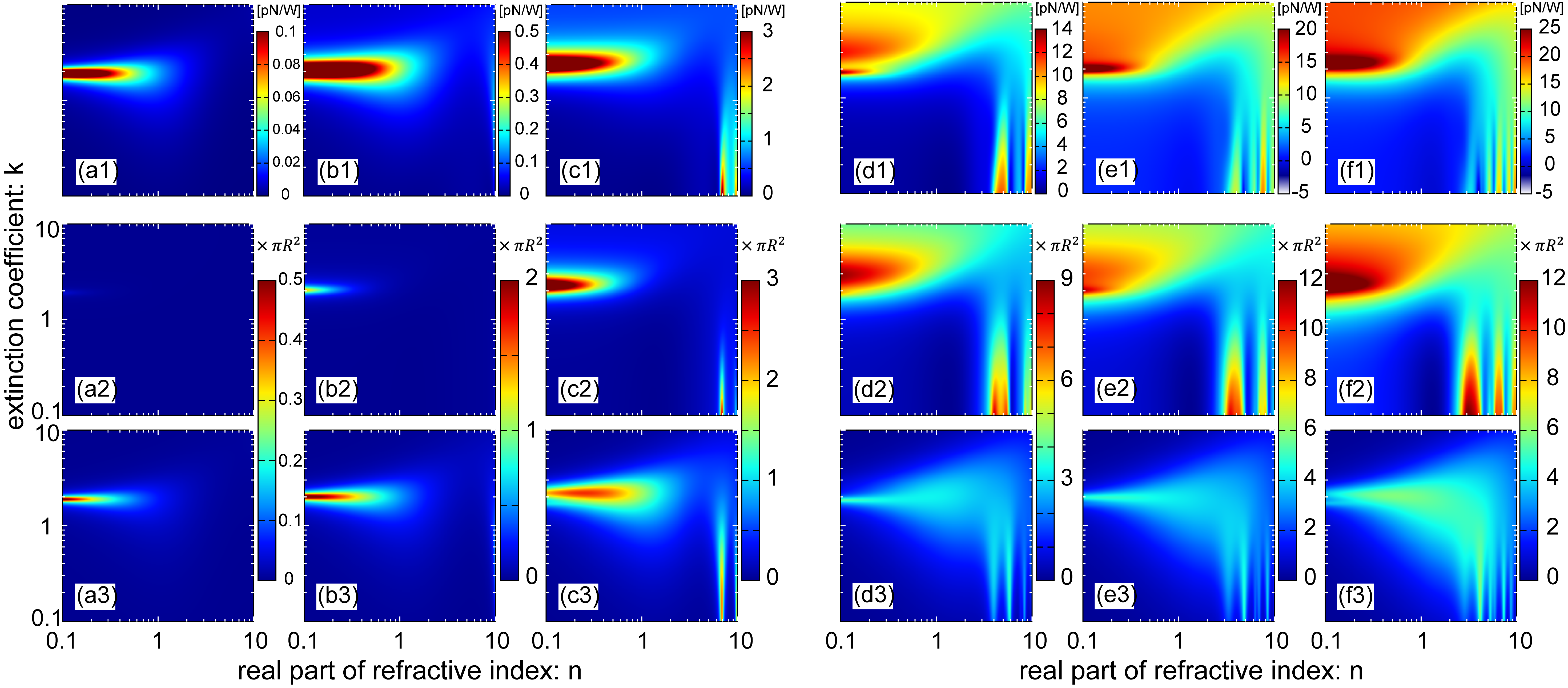}
\caption{
Indirect optical force and cross-sections in the plane of complex refractive index, $\tilde{n} = n + i\kappa$, for
$d=50\, \mathrm{nm}$(a) to $350\, \mathrm{nm}$(f), excluding $d = 200\, \mathrm{nm}$, when $N_{\rm p} = 7$;
the case of $d = 200\, \mathrm{nm}$ is shown in Fig.\ \ref{fig:NKplane} in the main text.
The optical force is projected in the angular direction and evaluated on one surrounding particle at
$r= \Delta =853.4\, \mathrm{nm}$.
Both cross-sections are plotted in a unit of $\pi R^2$ with $R=100\, \mathrm{nm}$.
}
\label{figS:FCsCa}
\end{figure*}

Figure \ref{figS:FCsCa} demonstrates the indirect optical force when $N_{\rm p} = 7$ and
the scattering and absorption cross-sections of a single particle for $d=50-350\, \mathrm{nm}$, excluding $d=200\, \mathrm{nm}$; the case of $d=200\, \mathrm{nm}$ is shown in the main text.
Note that the cross-sections in Fig.\ \ref{figS:FCsCa} are normalized by $\pi (100\, \mathrm{nm})^2$ for any $d$.
The optical force is evaluated when the surrounding particles are at $r= 853.4\, \mathrm{nm}$.
The force tends to increase with the particle size.
When $d<200\, \mathrm{nm}$, the optical force is much smaller than that of $d=200\, \mathrm{nm}$,
which corresponds to the depression of $C_{\rm sca}$.
At $d>200\, \mathrm{nm}$, the indirect optical force clearly increases with $d$,
whereas the scattering cross-section is increased only slightly.
Then, the scattered field intensity is not affected significantly by the diameter.
Meanwhile, the optical force increases according to the particle volume.
The stable binding distance $\Delta$ from the focal point is slightly affected by the particle size,
and one must evaluate the self-consistently.
However, we fixed the position of surrounding particles at $r = \Delta = 853.4\, \mathrm{nm}$
because the structures of force in the $n$-$\kappa$ plane shown in Fig.\ \ref{fig:NKplane} in the main text and
Figs.\ \ref{figS:FCsCa}(a1)--(f1) exhibit no significant change.

The absorption cross-section also depends on the diameter.
For large particles, the absorption cross-section tends to be relatively smaller than the scattering one.
In the $d<200\, \mathrm{nm}$ region, however, we can see a $C_{\rm sca} < C_{\rm abs}$ region in the plane.
From the figures of $d = 50$ and $100\, \mathrm{nm}$,
the behavior of indirect optical force in the $n$-$\kappa$ plane is similar to the absorption cross-section,
whereas the structures of optical force correspond to those of $C_{\rm sca}$ for $d \ge 150\, \mathrm{nm}$.
Then, we can say that the scattering cross-section is an essential factor for a large indirect optical force
when the scatterers have sufficient size.
However, for tiny particles, $C_{\rm sca}$ is too small and the absorption dominates the optical force.

The analysis of the SAM and OAM components of ``emitted'' light can also be examined.
However, discussion on the SAM and OAM with small or large particles is essentially equivalent to that the main text.

\section{Scattering of light with spin--orbit coupling}
\label{appen:SO}
\begin{figure}[t]
\includegraphics[width=0.9\linewidth]{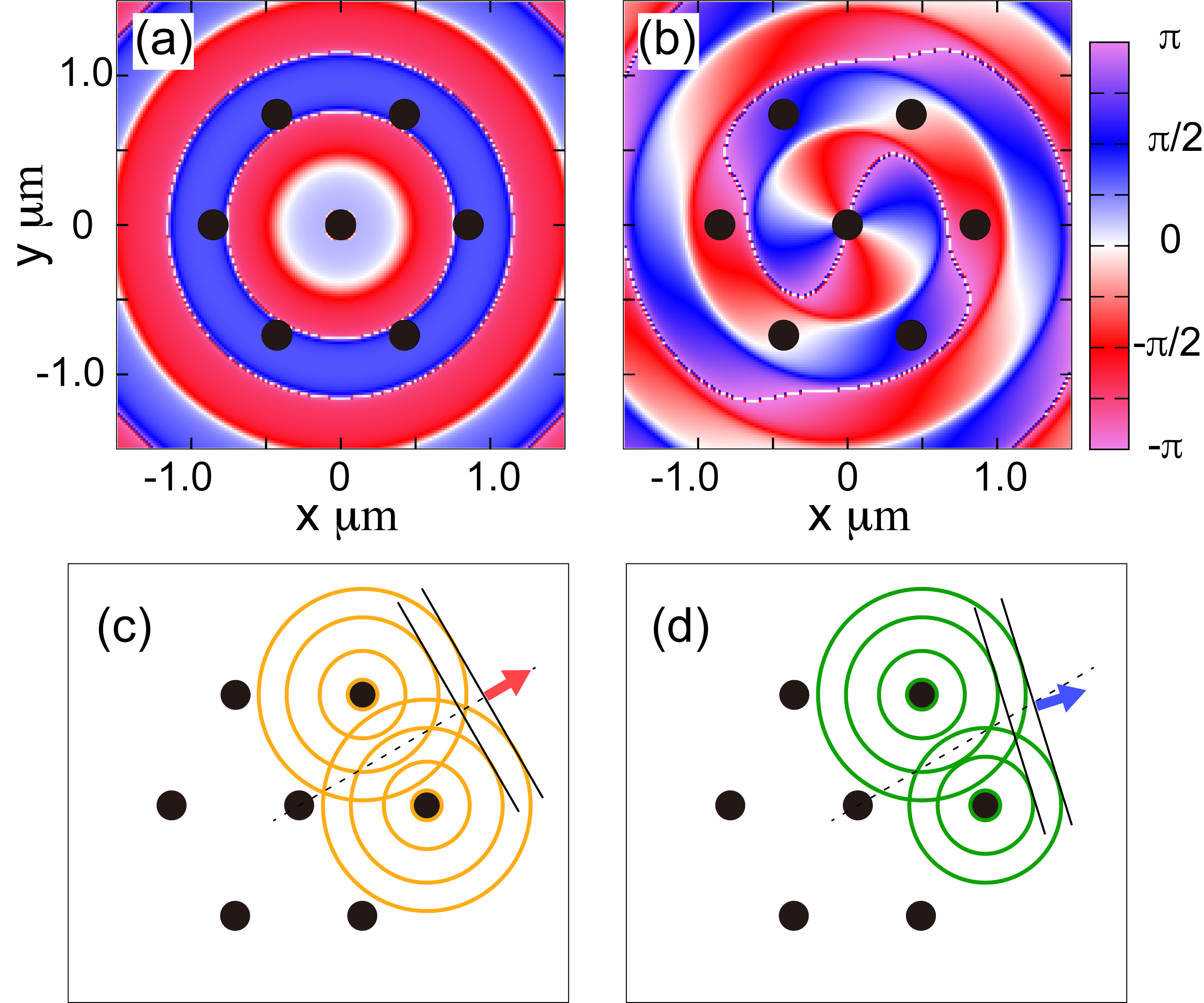}
\caption{
Phase profile of $l=0$ (a) and $l=2$ OAM (b) incident components.
Black circles show the particles with $d=200 \, \mathrm{nm}$ when $\Delta =853.4 \, \mathrm{nm}$.
The curve indicating a phase jump at $\pm \pi$ is artificial redundancy in (b).
(c,d) Schematic explanation of force by the light scattering when $l=0$ (c) and $l=2$ (d)
based on the phase of light and the Huygens--Fresnel principle.
}
\label{figS:OAM02}
\end{figure}

\addcontentsline{toc}{section}{Scattering of light with the spin--orbit coupling of light\dotfill}
The flip of SAM also explains the acceleration of surrounding particles.
Figure \ref{figS:OAM02} shows the mechanism schematically.
When $l=0$ OAM is scattered from the center particle, the scattered fields at six surrounding particles
have the same phase [see Fig.\ \ref{figS:OAM02}(a)].
Then, the Huygens--Fresnel principle from the surrounding particles results in light emission in the radial direction,
as shown in Fig.\ \ref{figS:OAM02}(c).
Meanwhile, for $l=2$ shown in Fig.\ \ref{figS:OAM02}(b), the scattered fields at the neighboring two surrounding particles
have $4\pi /6$ phase differences.
This phase difference causes a tilt of wavefront of emitted light, as in Fig.\ \ref{figS:OAM02}(d).
This tilt of ``light emission'' causes a scattering force and accelerates the surrounding particles.

If the ``SO coupling'' of light at a single particle is enlarged, the ratio of $l=2$ scattering from
the center particle against $l=0$ scattering increases.
Therefore, by this schematic understanding, stronger SO coupling results in stronger indirect optical force,
which is reasonable from the similarity between the profiles of Figs.\ \ref{fig:NKplane}(a) and (c) in the main text.

\section{Description of movie}
The movie in this  shows a numerical simulation of the dynamics of gold particles in a water solvent.
The parameters of the simulation are the same as those of Fig.\ \ref{fig:OptBind}(e) in the main text.
In this movie, we assume a slight charge on the particles, which might exist experimentally.
However, the presence of charge on the particles only slightly changes the particle distance 
and is not essential for the dynamics.


\section*{ACKNOWLEDGMENT}

The authors thank Prof.\ H.\ Masuhara, Dr.\ T.\ Kudo, and Z.-H.\ Huang for their fruitful discussions on their experimental results.
The authors acknowledge for all member of Collective Optofluidic Dynamics of Nanoparticles meeting organized by Prof.\ Masuhara.
This work was supported by JSPS KAKENHI Grant Number 16K21732 in Scientific Research on Innovative Areas ``Nano-Material Optical-Manipulation''.
T.Y.\ was supported by JSPS KAKENHI Grant Number 18K13484 and H.I.\ was  supported by JSPS KAKENHI Grant Number 16H06504 and 18H01151.

\bibliography{acs-YTTYHI}
\bibliographystyle{unsrt}

\end{document}